\documentclass[conference,final,twocolumn]{IEEEtran}
\usepackage{amsmath,amssymb,amsfonts,mathrsfs,bm}
\usepackage{amstext}
\usepackage{upgreek}
\usepackage{multicol}
\usepackage{graphicx}

\usepackage{paralist}

\usepackage[numbers,sort&compress]{natbib}
\usepackage{booktabs}
\usepackage{multirow}
\usepackage{subfigure}

\usepackage{algorithm}
\usepackage{algorithmic}

\IEEEoverridecommandlockouts

\begin{document}
\title{Blind Receive Beamforming for Autonomous Grant-Free High-Overloading Multiple Access}
\author{
\IEEEauthorblockN{{Zhifeng Yuan, Weimin Li, Yuzhou Hu, Xun Yang, Hong Tang, Jianqiang Dai}}
\IEEEauthorblockA{ZTE Corporation, Shenzhen, China\\
Email: \{yuan.zhifeng, li.weimin6, hu.yuzhou, yang.xun3, tang.hong3, dai.jianqiang\}@zte.com.cn}

%
}
\maketitle
\begin{abstract}
  Massive number of internet of things (IoT) devices are expected to simultaneously connect to the mMTC and beyond future generations of wireless network, posing severe challenge to aspects such as RACH procedure, user equipment detection and channel estimation. Although spatial combining has provided significant gains in conventional grant-based transmission, this technique is stuck in dilemma when it comes to autonomous grant-free transmission tailored for IoT use cases. To address this, blind spatial combining and its incorporation in the data-only MUD are elaborated in this paper answering to both the academic and industry's concern in the overloading potential of autonomous grant-free (AGF) transmission. Blind spatial combining  could be interpreted as blind receive beamforming heuristically. Simulation results show that the blind spatial combining enhanced data-only MUD performance for AGF transmission is rather impressive.
\end{abstract}

\begin{IEEEkeywords}
autonomous grant-free, high-overloading transmission, collision, spatial combining, blind receive beamforming, combining vectors, blind multi-user detection
\end{IEEEkeywords}

\section{Introduction}
Massive Machine-type communication (mMTC) is widely anticipated to be a very important scenario in the future generations of wireless network \cite{3GPP.802,Yuany.NOT,LDai.NomaSol}. In this scenario, typical use cases include the Internet of Things (IoT) applications such as pervasive ubiquitous ehealth. These applications require simultaneous connections of a massive number of low-data-rate devices \cite{bavand2017maximum,bavand2017user}. Revolutionary Autonomous Grant-Free (AGF) access \cite{axiv,zYuanvtc18,Yuan.BMU} allowing UE transmission directly from idle state can extremely simplify the transmitter design and is tailored for IoT systems in mMTC and beyond scenarios thanks to the merits that come with this access mode \cite{axiv,zYuanvtc18}. High overloading (HOL) capacity enables the simultaneous transmissions of massive sporadic small packets with reasonable bandwidth. Thus AGF-HOL transmission combining these two desirable features could be very valuable for economic deployment of mMTC and beyond wireless systems \cite{axiv,zYuanvtc18}. However the multi-user detection (MUD) achieving AGF-HOL is indeed very challenging in the sense that not only great blind detection efforts are needed as the access information is unknown to the receiver, but more intractably the critical multiple-access (MA) signature (including spreading code, scrambling, modulation pattern and reference signal etc) collision problem would worsen rapidly with the AGF loading and cause severe system performance degradation. Clear industry interest in this regard has been demonstrated as the consensus has been reached in the 3GPP standardization organization to study the MA signature collision impact in the non-orthogonal multiple access (NoMA) study item \cite{ZTErSINoMA}. Take reference signal (RS) collision as an example, if two UEs have selected the same RS, not only would the UE detection be affected but the channel estimation (CE) could be so deviated that the equalization in the context of high-overloading would be deteriorated even for the detected UE and therefore the detected UE would be highly unlikely to be successfully decoded whatever so-called advanced receiver such as MPA/EPA/ESE receiver is employed. The impact of channel estimation accuracy on the  performance of the relevant receivers have been investigated in the literature \cite{zhou2007iterative,sergienko2017spectral,hammarberg2012channel}.   To address the above critical issues, the paper \cite{axiv} provided four potential discriminations among multiplexed UEs committing AGF transmissions and furthermore introduced a state-of-art blind MUD which can achieve AGF-HOL even at single receive antenna case. Concretely the blind MUD in \cite{axiv,zYuanvtc18} has exploited only three discriminations except the spatial discrimination to achieve over 300\% AGF overloading. Furthermore simulation results in \cite{Yuan.BMU,ZTE.perf} have shown that remarkable 600\% and 1000\% AGF overloading can be achieved by further exploiting the spatial discrimination offered by two and four receive antennas respectively. It can be seen that even with truly blind MUD the system loading of AGF can also achieve nearly linear increase with the number of receive antenna as long as the spatial discriminations can be fully utilized. However how to utilize the multiple receive antennas effectively in AGF-HOL blind MUD is a crucial problem and has not yet been revealed in \cite{axiv,Yuan.BMU,ZTE.perf}, which will be the task of this paper. In this paper, some novel ideas to achieve AGF-HOL with multiple receive antennas are discussed first and then a concrete blind MUD receiver capable of combining the powerful spatial discrimination with the other three discriminations to realize AGF-HOL capacity in an implementation friendly manner is proposed. AGF transmission without RS, a.k.a. data-only AGF, is emphasized in this paper because it can provide one more discrimination than AGF transmission with RS, as analyzed in \cite{axiv}. However the novel ideas exploiting spatial discrimination can also significantly improve the system loading for the AGF transmission with RS (preamble included) or even some specific grant-based transmission.

The paper is organized as follows. In Section II, the problems of applying conventional spatial approach to exploit multiple receive antenna gains in AGF-HOL are presented first and then novel blind spatial approach addressing the difficulties is proposed. To reduce the complexity of the new idea, blind activity detection methods borrowed and extended from that in \cite{zYuanvtc18} are introduced in Section III, which can significantly narrow down the blind detection range. In Section IV, performance evaluations comparing the blind spatial combining enhanced data-only MUD with ideal MMSE-SIC are provided. Performance evaluations by means of conventional spatial approaches are also attached for reference. Section V concludes the entire paper.

\section{ Conventional Spatial Combining Techniques versus Blind Spatial Combining}
\begin{figure}[!htb]
    \centering
     \includegraphics[width=0.5\textwidth]{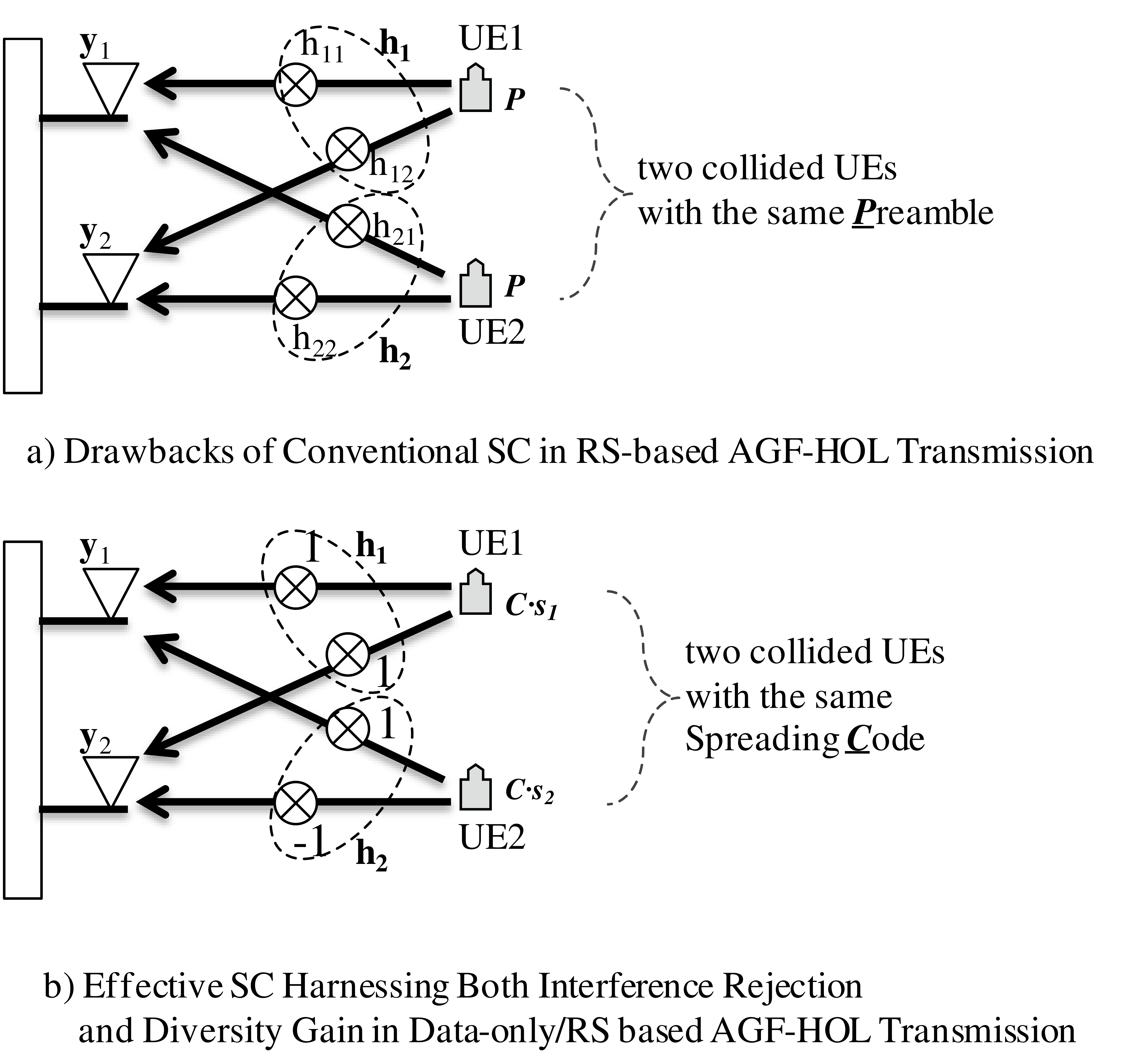}\\
	\caption{Conventional SC vs BSC in AGF-HOL Transmission}\label{col2}
\end{figure}

Suitable spatial combining (SC) of the signals received by multiple antennas can be exploited to harness the spatial gains such as diversity gain and interference rejection gain, especially for cell-edge users in an interference-limited system \cite{yang2011receive}. Conventional spatial combining (CSC) such as Maximum Ratio Combining (MRC) and Zero Forcing Combining (ZFC) require the knowledge of spatial channels per antenna to be known prior to the combining, which means channel estimation needs to be performed at each receive antenna separately. In grant-based or semi-persistent scheduling (SPS) UL transmission, antenna-wise channel estimation could be performed at relatively higher accuracy at the expense of more overhead dedicated for enhanced RS design.
However in AGF transmission, where UEs randomly select the MA signatures such as the RS and spreading code, incontrollable collision of RS and spreading code degrades significantly the antenna-wise channel estimation for RS-based AGF and data-only AGF respectively as the overloading grows. Fig. \ref{col2} a) puts forward an intuitive example showing how the antenna-wise channel estimation of preamble-based AGF is affected by preamble collision. Two receive antennas in the base station (BS) and one transmit antenna in the user equipment (UE) are assumed. UE1 and UE2 select the same preamble {\bf P} in preamble-based AGF access in Fig. \ref{col2} a) and they are dubbed as collided UEs. The spatial channel vector of UE $k$ is denoted as ${\bf h}_k=[h_{k,1},h_{k,2}]$. The receive signals associated with the preamble in antenna 1 and 2 are ${\bf y}_1$, ${\bf y}_2$ respectively, while ${\bf n}_1$ and ${\bf n}_2$ denote the additive White Gaussian Noise (AWGN) at the receive antenna respectively. The transceiver equation could be elaborated as follows,
\begin{equation}
\begin{aligned}
\label{tEqu}
{\bf y}_1 = {\bf P} h_{11}+{\bf P}h_{21}+ {\bf n}_1 \\
{\bf y}_2 ={\bf P}h_{12}+{\bf P}h_{22}+ {\bf n}_2
\end{aligned}
\end{equation}

\begin{figure}[!htb]
    \centering
     \includegraphics[width=0.5\textwidth]{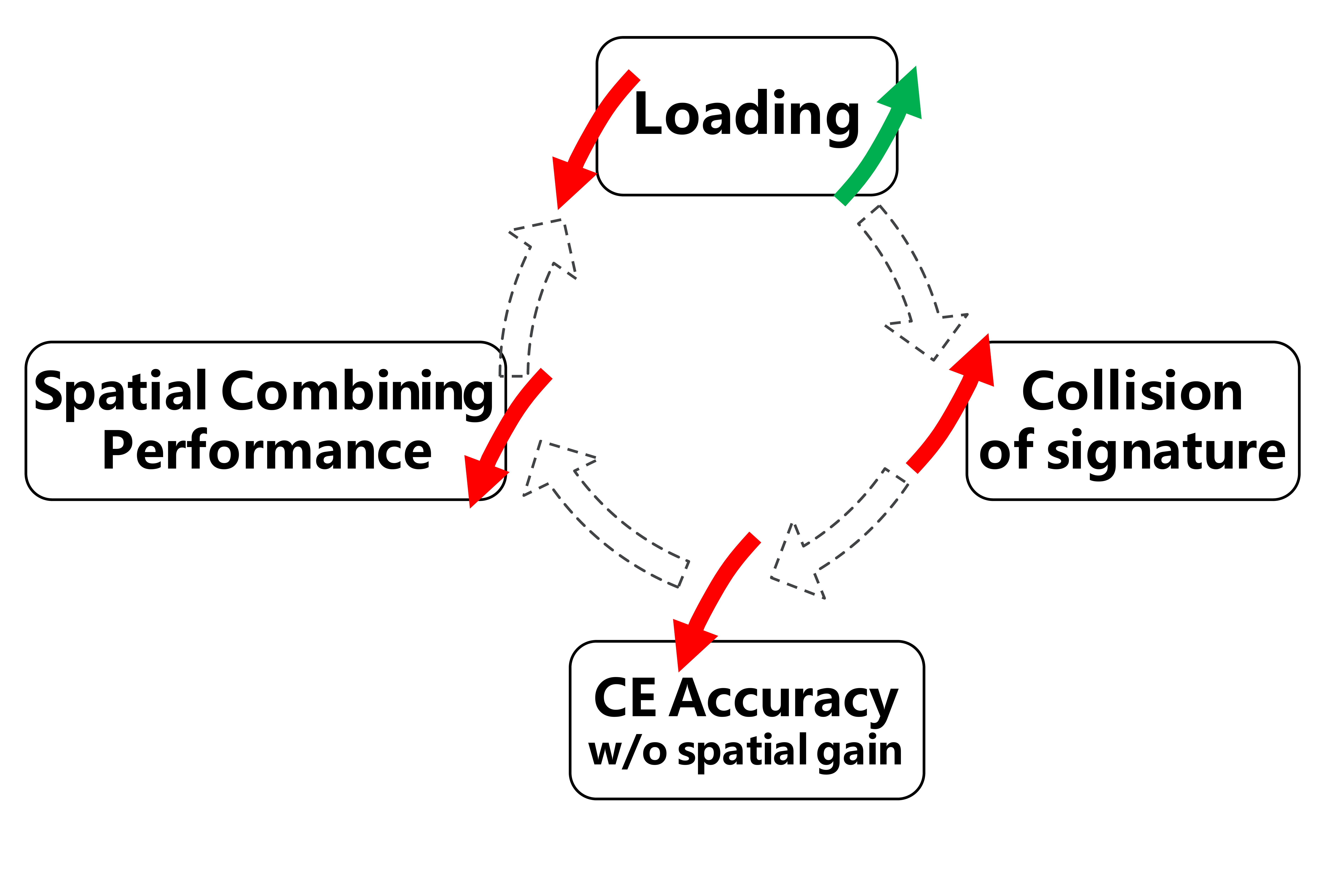}\\
	\caption{Dilemma Faced by Conventional SC}\label{Dil}
\end{figure}

According to \eqref{tEqu}, conventional preamble-based channel estimated antenna-wise through ${\bf y}_1$ and ${\bf y}_2$ is $h_{11} + h_{21}$ and $h_{12}+h_{22}$ respectively, resulting in an estimated spatial channel vector of $\hat {\bf h} = {\bf h}_1 + {\bf h}_2$ for both UE1 and UE2. It could therefore be inferred that conventional SC such as MRC or ZFC based on the estimated $\hat {\bf h}$ could result in combined SNR below expectation and subsequent degraded demodulation performance. It should be emphasized that the collision situation depicted in Fig. \ref{col2} a) appears unavoidable and is indeed the performance bottleneck in the AGF-HOL transmission.

As to the data-only AGF transceiver, although data only blind MUD could provide more discriminations to handle signature collision than preamble-based AGF counterpart (i.e. it could still provide acceptable channel estimation for the stronger one of the signature-collided UEs, and then decode and subtract the signal of the stronger UE), the performance of the blind MUD approach such as Partition-Matching (PM) blind channel estimation method \cite{axiv,zYuanvtc18} would degrade as well if the collision probability exceeds a certain level as the loading increases. Thus if data-only MUD including PM method based blind channel estimation is merely applied at each receive antenna separately, the loading capacity cannot exceed the level of single receive antenna significantly, since the spatial domain discrimination from multiple receive antennas could hardly be effectively exploited.

\begin{figure}[!htb]
    \centering
     \includegraphics[width=0.5\textwidth]{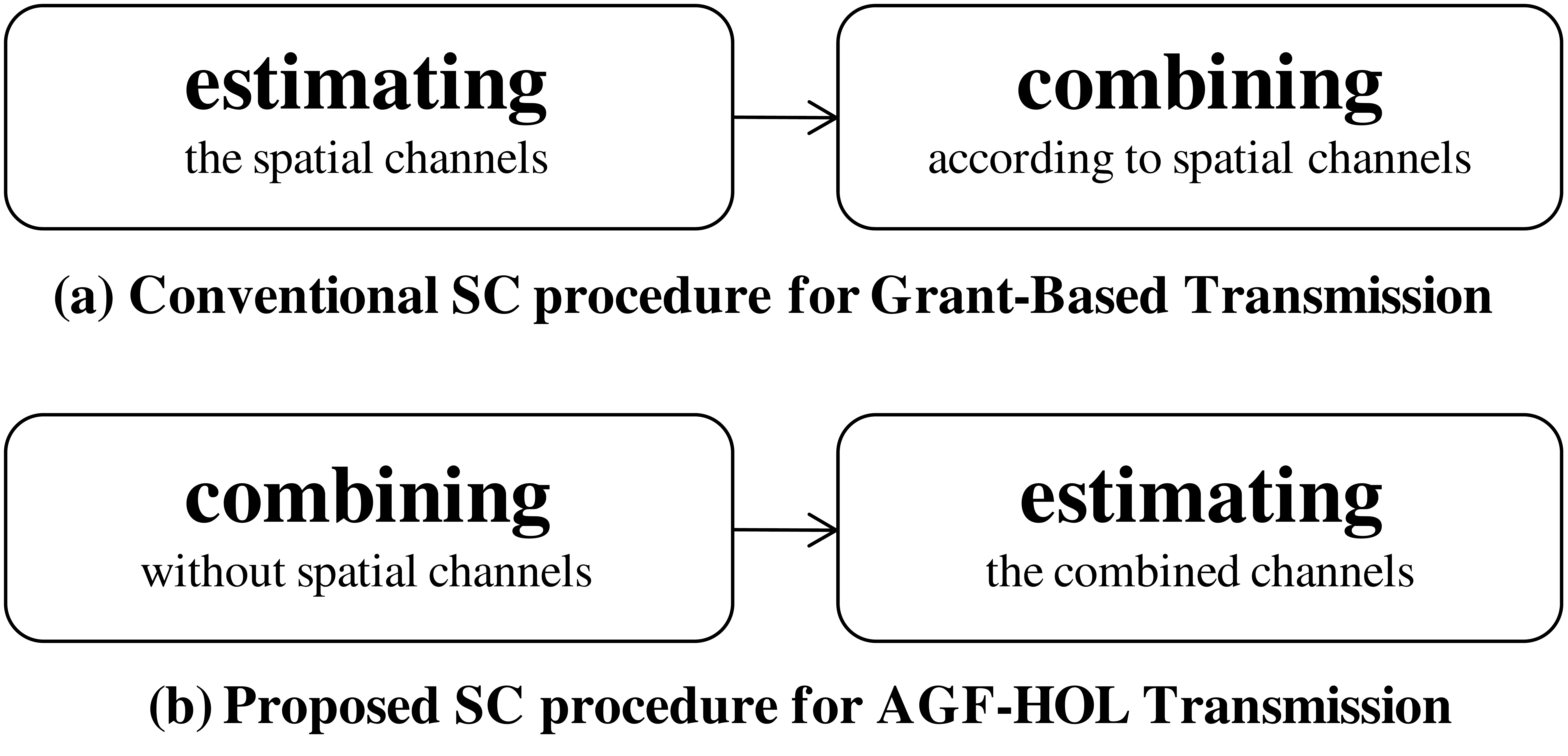}\\
	\caption{Protocols for BSC and Conventional SC}\label{BSCvsCSC}
\end{figure}

The dilemma faced by CSC applied to the AGF-HOL transmission could be further summarized in Fig. \ref{Dil}. As the loading increases, the probability of preamble/spreading code collision increases rapidly and therefore the channel estimation accuracy performed antenna-wise would be deteriorated and SC based on the channel estimation would not bring about the expected SNR improvement and performance gain. The key point of this dilemma is the conventional spatial procedure ¡®estimating first and combining afterwards¡¯, since the estimating accuracy of spatial channel without spatial combining gain would worsen as a result of the high signature collision level in the AGF-HOL. In order to break the dilemma, a new idea by just reversing the order of the conventional procedure, resulting in ¡®combining first and estimating afterwards¡¯, seems reasonable since it could harness the spatial combining gain as early as possible and then the channel estimation afterwards would work in a collision-alleviated fashion with increased SINR, leading to better demodulation performance following the more accurate channel estimation. This reversing can be outlined as in Fig. \ref{BSCvsCSC}. It should be noted that performing effective SC without the UEs¡¯ spatial channels is the key action in our blind MUD for AGF-HOL.

But how to perform effective SC in high-collision situation without the knowledge of UEs¡¯ spatial channels? This question can be answered heuristically by a simple example associated data-only AGF shown in Fig. 1 b) where the two UEs have selected the same spreading code {\bf C} composing of spread elements $c$ and the specific spatial channel vectors of UE1 and UE2 are ${\bf h}_1=[1,1]$ and ${\bf h}_2=[1,-1]$ respectively. Below is the chip-wise scalar form transceiver equation,
\begin{equation}
\label{scaEqu}
\begin{aligned}
y_1 =  cs_1+cs_2+n_1 \\
y_2 = cs_1-cs_2+n_2 \\
\end{aligned}
\end{equation}

Obviously, it's a severe equal-power collision in each receive antenna and would increase the difficulty of the MUD if related MUD techniques are merely applied in each receive antenna separately. However if two pre-defined linear combinations such as ${\bf z}_1={\bf y}_1+{\bf y}_2$, ${\bf z}_2={\bf y}_1-{\bf y}_2$ are used firstly, then the chip-wise scalar form equation after these two combinations is

\begin{equation}
\label{comb}
\begin{aligned}
z_1 = 2cs_1 + n_1 + n_2 \\
z_2 = 2cs_2 + n_1 - n_2
\end{aligned}
\end{equation}

Then $z_k$ contains only UE $k$'s signal with AWGN. Thus the severe equal-power collision is separated perfectly and a much higher SINR is achieved. The pre-defined combining can work without the knowledge of the UEs¡¯ spatial channel vectors, thus can be viewed as ¡®blind¡¯ spatial combining (BSC). Linear combining of the receive symbol streams can also be implemented by inner product of combining vector and the receive symbol vectors. Concretely the above two linear combining actions are equivalent to these two inner products: ${\bf z}_1={\bf v}_1{\bf Y}$, ${\bf z}_2={\bf v}_2{\bf Y}$, where ${\bf v}_1=[1,1]$ and ${\bf v}_2=[1,-1]$ are the pre-defined combining vectors and ${\bf Y} = [y_1, y_2]^{t}$ is the receive signal vectors from the two receive antennas. All these are vectors in  ${\bf C}^{m}$ space, where $m$ is the number of the receive antennas and $m$ equals to 2 in this example.

\begin{table}
\begin{center}
\caption{ $\mathcal{V}_4$ consisting of 24 combining vectors. }
\begin{tabular}{l*{2}{c}r}

Index              & Vector & Index & Vector  \\
\hline
1 & [1,1,0,0] & 13 & [1,i,0,0]  \\
2 & [1,-1,0,0] & 14 & [1,-i,0,0] \\
3 & [1,0,1,0] & 15 & [1,0,i,0]  \\
4 & [1,0,-1,0] & 16 & [1,0,-i,0]  \\
5 & [1,0,0,1] & 17 & [1,0,0,i]  \\
6 & [1,0,0,-1] & 18 & [1,0,0,-i] \\
7 & [0,1,1,0] & 19 & [0,1,i,0]  \\
8 & [0,1,-1,0] & 20 & [0,1,-i,0]  \\
9 & [0,1,0,1] &21 &[0,1,0,i] \\
10& [0,1,0,-1] &22 &[0,1,0,-i]\\
11& [0,0,1,1] &23 &[0,0,1,i]\\
12& [0,0,1,-1]&24 &[0,0,1,-i]
\end{tabular}
\end{center}
\end{table}

\begin{table}
\begin{center}
\caption{ $\mathcal{V}_4$ consisting of 16 combining vectors. }
\begin{tabular}{l*{2}{c}r}

Index              & Vector & Index & Vector  \\
\hline
1 & [1,1,1,1] & 9 & [1,-1,-i,-i]  \\
2 & [1,-1,1,-1] & 10 & [1,1,-i,i] \\
3 & [1,1,-1,-1] & 11 & [1,-1,i,i]  \\
4 & [1,-1,-1,1] & 12 & [1,1,i,-i]  \\
5 & [1,-i,-i,1] & 13 & [1,i,-1,i]  \\
6 & [1,i,i,-1] & 14 & [1,-i,-1,-i] \\
7 & [1,-i,-i,-1] & 15 & [1,i,1,-i]  \\
8 & [1,i,-i,1] & 16 & [1,-i,1,i]  \\

\end{tabular}
\end{center}
\end{table}

\begin{figure}[!htb]
    \centering
     \includegraphics[width=0.5\textwidth]{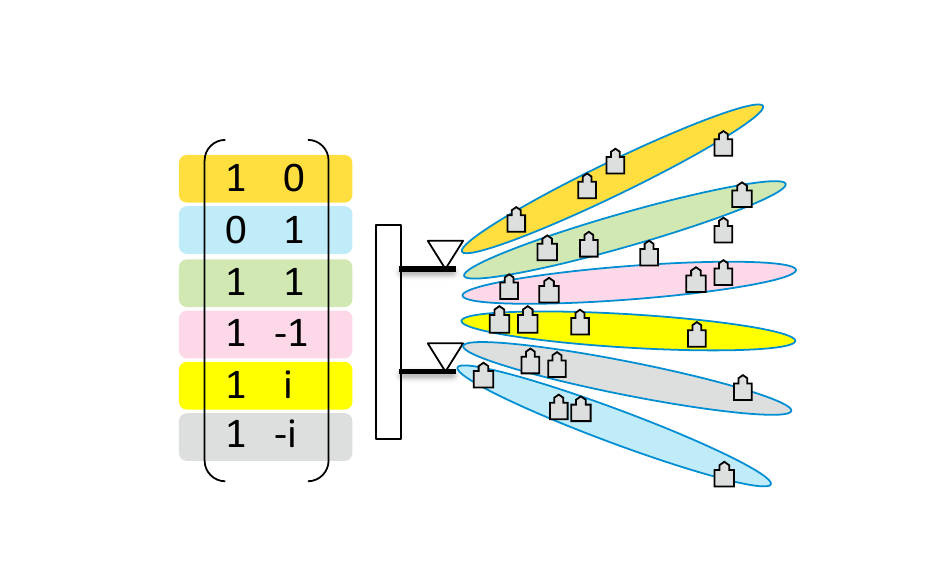}\\
	\caption{BSC from the Perspective of Receive Beamforming}\label{BSCinRBF}
\end{figure}

\begin{figure}[!htb]
    \centering
     \includegraphics[width=0.5\textwidth]{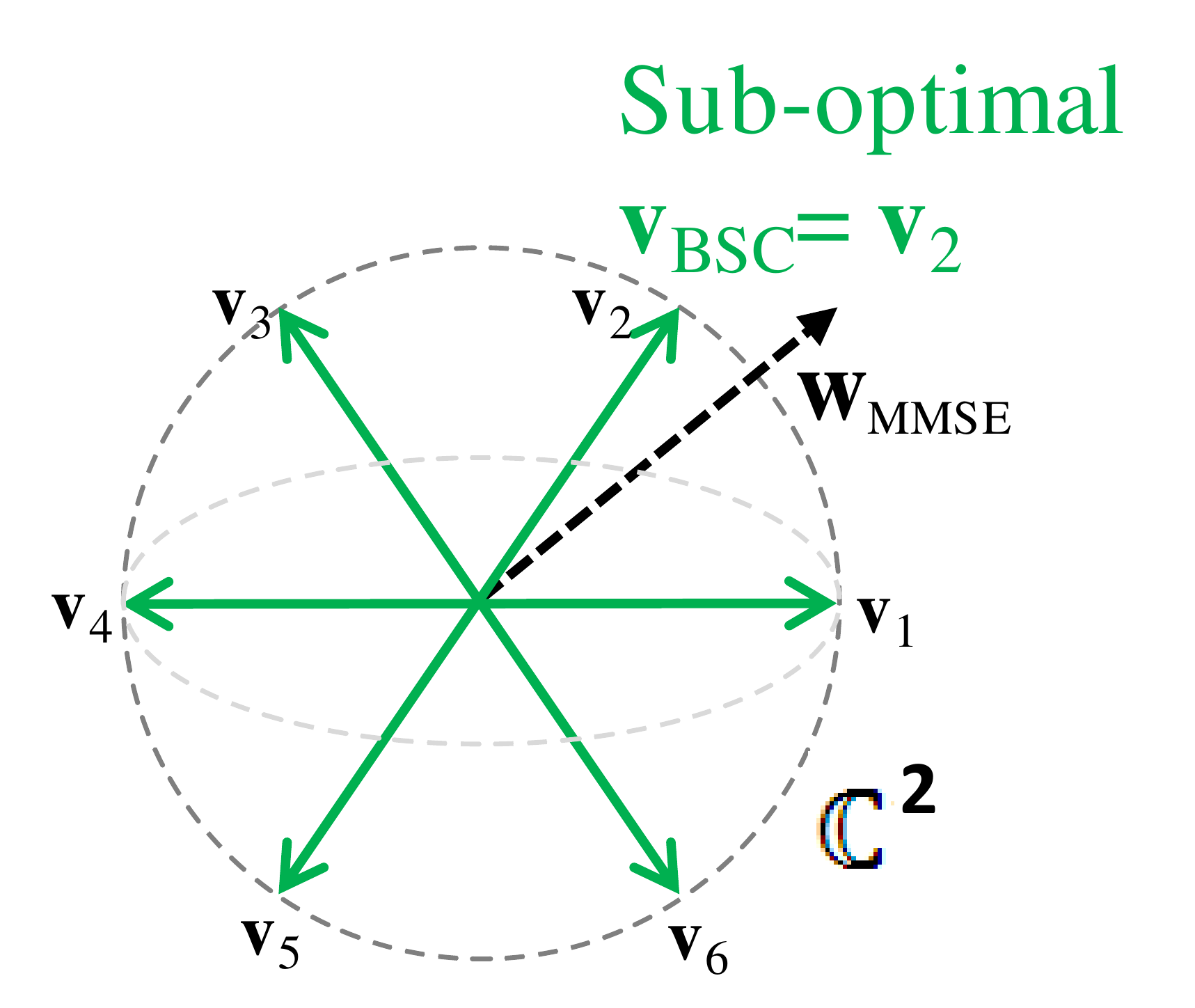}\\
	\caption{Sub-optimal BSC Vector Set vs Optimal MMSE Combining vector}\label{BSCVSet}
\end{figure}

The design of the pre-defined combining weights ${\bf v}$ are critical for BSC. Although the optimal linear combining weights can not be obtained due to the probable inaccurate channel estimation in AGF, we could still strive for a set of pre-defined  $m_v$ combining vectors  for $m$ receive antennas considering the trade-off between performance and complexity. Intuitively speaking, a uniform division of the $\mathbf{C}^m$ would be a great candidate for the set $\mathcal{V}_m$ of the $m$ receive antenna BSC combining vectors ${\bf v}_1, {\bf v}_2,...,{\bf v}_{m_{v}}$ and increasing the cardinality $m_v$ of the $\mathcal{V}$ would lead to both superior AGF MUD performance and complexity. In case of two receive antennas, we propose $\mathcal{V}_2$ of size 6 as follows, ${\bf v}_1 = [1,0]^{t}, {\bf v}_2 = [0,1]^{t}, {\bf v}_3=[1,1]^{t}, {\bf v}_4=[1,-1]^{t}, {\bf v}_5=[1,i]^{t}, {\bf v}_6=[1,-i]^{t}$. Note that the vectors should be normalized. Not only do these vectors achieve a quasi-uniform partition of the $\mathbf{C}^2$ space but the linear combination operations could also be simplified into additions from multiplications based on the algebra property of complex numbers. Actually the nearly $600\%$ AGF-HOL transmission overloading is achieved employing $\mathcal{V}_2$ \cite{Yuan.BMU,ZTE.perf}. $\mathcal{V}_4$ consisting of 24 and 16 combining vectors for 4 receive antennas are attached in Table I/II above, which are also designed considering the performance and complexity trade-off as $\mathcal{V}_2$ for 2 receive antennas.  $1000\%$ overloading has been achieved employing the 24 combining vectors.
Another heuristic interpretation of the BSC gain comes from the perspective of receive beamforming. Six receive beams could be constructed utilizing the 6 combining vectors in $\mathcal{V}_2$. The channel vectors distribute in a random  fashion in the $\mathbf{C}^2$ space as illustrated in Fig. \ref{BSCinRBF}. Each beam could align some UEs while cancel the signals from other UEs. Diversity and interference rejection gain could be obtained to a certain extent in the statistical sense. In data-only MUD, each of the combining vectors in $\mathcal{V}_{m}$ would be employed to perform BSC and the combined data stream would go through the blind MUD operations revealed in \cite{axiv,zYuanvtc18}. The hard-decision SINR (HD SINR) after blind equalization would be sorted for the equalized data streams generated from all the $m_v$ combining vectors and the UEs possessing the largest post-SINR (the SINR of the equalized data stream)  are likely to possess both relatively better channel gain and matching combining vectors. Take a more practical point of view, the gap between the sub-optimal pre-defined BSC combining vector and the optimal combining vector in Fig. \ref{BSCVSet} could be viewed as vector quantization error \cite{gray1990vector}. So long as this vector quantization error doesn't degrade the post-SINR below the threshold, the successfully decoded and reconstructed symbols obtained from the decoded data could still be leveraged to perform an accurate channel estimation and fine cancellation of the stream from the aggregate signal \cite{axiv, zYuanvtc18}. In this way the data-only MUD and successive interference cancellation (SIC) could continue. Simulation results demonstrate indeed that at least in the data-only AGF case the quantization error doesn't lead to much performance degradation compared with the upper bound where perfect channel estimation is assumed. In addition, there is some margin in the MCS for the randomness of AGF-HOL access, i.e. low-order modulation scheme such as BPSK coupled with low effective code rate, thanks to which the decoding threshold is relatively low. It's worth mentioning that the low-data-rate power efficient one-dimensional BPSK modulated signals are of interest to reliably support emerging IoT systems consisting of massive numbers of low-data-rate devices. Moreover, BPSK is a commonly employed transmission mode in widely deployed wireless systems such as IEEE 802.11a,n,ac, especially when SNR at the operating point is low \cite{abdallah2014rate}.

\begin{figure*}[!htb]
    \centering
     \includegraphics[width=\textwidth, scale = 2]{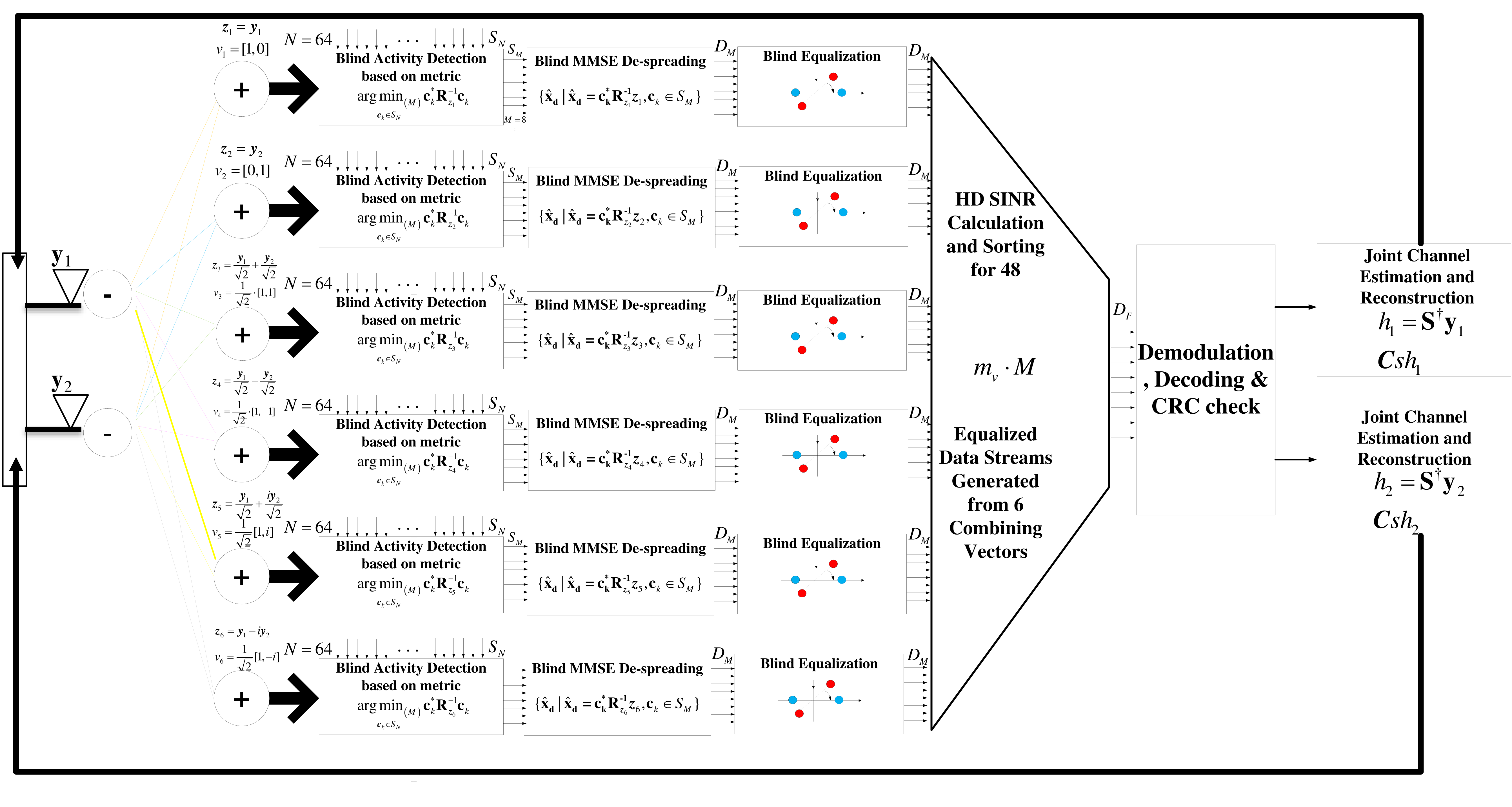}\\
	\caption{Data-only AGF MUD with BSC}\label{dataOnlyMUD}
\end{figure*}

\section{Architecture of Blind MUD with BSC for AGF }
The spatial domain discrimination enabled by BSC could benefit AGF transmission both with and without RS. These two types of AGF transmissions behave nevertheless differently as the loading and collision probability increases. With the employment of BSC, although the preamble/spreading code collision issues could be alleviated in the sense that inter-beam collision could be handled taking into account the beamforming interpretation of the BSC. However intra-beam preamble collision caused combined channel estimation of all the collided UEs difficulty still resides. This causes a severe performance degradation not only in UE detection but in decorrelation operations for the detected UEs. On the other hand, data-only AGF could exploit discriminations both in the power domain and constellation domain relying on PM method due to the fact that the transmit data independence could to a certain extent compensate the spreading code collision induced degradation. The above analysis could account for the diversified performance between the RS-based and data-only AGF transmission. This section therefore concentrates on the more promising data-only MUD with BSC enhancements for AGF-HOL transmission.

The BSC module is a key component in our data-only MUD, however after BSC, i.e. after $m_v$ pre-defined combining, the receiver still has no knowledge of which UEs are in each of the $m_v$ combined streams, what are their channel gain factors, and which spreading codes are selected by  them etc. Therefore state-of-art blind detection procedures relying on the data symbols are needed after the BSC. Furthermore, the incorporation of blind receive beamforming and blind detection also implies that the data-only blind detections should be performed exhaustively over all the $m_v$ combined streams, whose computational complexity seems expensive at first glance. Thus low complexity blind activity detection techniques are needed following the BSC module. The following bullets outline the whole picture as illustrated in Fig. \ref{dataOnlyMUD} of a low-complexity data-only blind MUD revealed in detail in \cite{axiv,zYuanvtc18} in conjunction with BSC mechanism for multiple receive antenna case.

\begin{itemize}
    \item blind activity detection based on MMSE metric for each combining vector
    \item blind MMSE de-spreading using the detected codes for each combining vector
    \item blind channel estimation/equalization employing the partition matching (PM) method for each combining vector
    \item decoding a certain number of equalized streams according to hard-decision SINR sorting among the equalized data streams generated from all combining vectors
    \item refining spatial channel estimations via the reconstructed symbols based on which the decoded streams are cancelled.
\end{itemize}

Speaking of blind MUD, one may first be left with the impression that the computation complexity is alarmingly high. Indeed without the application of the state-of-art blind activity detection based on MMSE metric, the blind despreading and equalization efforts would be at the order $N$ i.e. cardinality of the spreading code set $S_N$, which is a relatively large number. The blind MMSE metric in the context of multiple receive antennas is elaborated as follows,
\begin{equation}
\mathrm{argmin}_{(M)} {\bf c}_k^{*} R_{\bf z}^{-1}{\bf c}_k, {\bf c}_k \in S_N
\end{equation}
where ${\bf c}_k \in \mathbf{C}^{L \times 1}$ is the $k$-th spreading code of length $L$ in $S_N$, ${\bf c}_k^{*}$ is the conjugate of ${\bf c}_k$, and $R_{\bf z}$, a square matrix of dimension $L$, represents the correlation matrix of the combined spreading symbol ${\bf z}$. $\mathrm{argmin}_{(M)}$ finds the $M$ arguments within $S_N$ whose  ${\bf c}_k^{*} R_{\bf z}^{-1}c_k$ are smaller than that of others. Then blind despreading and the following blind equalization operations are only needed for $M$ detected codes instead of all the $N$ codes. The simulation results in \cite{axiv,zYuanvtc18} show that de-spreading and equalization can be performed for $M=8$ detected spreading codes instead of all the $N=64$ codes without much performance degradation in the single receive antenna case, bringing about 7/8 reduction of blind detection efforts, which is very important to efficient blind MUD. Motivated by this observation, blind activity detection based on other metrics deserves further research.

It is worth mentioning that feeding a certain number of equalized data streams possessing the highest hard-decision SINR to the FEC decoder is an extension of blind activity detection with the merit of reduced decoding efforts. The underpinning logic is that the active UEs with better channel gains and having taken better advantage of the BSC gains are more probable to possess higher HD SINR and therefore more likely to be successfully decoded. In Fig. \ref{dataOnlyMUD}, $F=8$ data streams are picked out and fed to FEC decoder based on HD SINR metric complying with the simulation setting in section IV. As illustrated in Fig. \ref{dataOnlyMUD}, the mechanism adopts a gradual reduction blind activity detection strategy progressively narrowing down the range of active UEs step by step . The hierarchical blind activity detection mechanism is summarized as follows, among which the CRC check is the most accurate component while the MMSE metric and the HD SINR metric could save blind despreading plus equalization efforts and decoding efforts respectively.
\begin{itemize}
  \item blind activity detection based on MMSE metric : $m_v \times M$ streams detected
  \item HD-SINR sorting and decoding $F$ data streams possessing highest HD SINR
  \item CRC check for the $F$ decoded data streams
\end{itemize}

\section{Performance Evaluation}
This section evaluates the performance of the data-only MUD architecture with BSC employing the MUSA sequence set\cite{axiv} as an example in high-overloading AGF scenario via link level simulations, common simulation assumptions are elaborated as follows. Transport block size is set at 84 bits including CRC. Turbo code at 1/2 code rate and BPSK modulation are assumed. The spreading code of each UE is assumed to be randomly selected from the length-4 sequence pool of size 64 in \cite{axiv}, and each spread-unit occupies 4 consecutive CP-OFDM symbols on one subcarrier. A total of 4 LTE resource blocks (RBs) consisting of 668 resource elements (REs) are employed to carry the spread symbols. Single transmit antenna and two receive antennas are assumed. SNR is defined at data RE level for the spreading data while the SNR given in \cite{axiv} is the transmit symbol level SNR with a spreading gain. Flat channel fading is assumed. It should be noted that with the variations of the size of blind detected sequence pool $S_M$ or the size of blind equalized data streams delivered to decoding pool $S_F$, performance fluctuations could be expected. As $M$ and $F$ increases, the blind despreading plus equalization burden increases and the decoding burden increases respectively, and the block error rate (BLER) performance is supposed to be better. In the following simulations, the numbers  are set to be 8 and 8/16/48 respectively reflecting the performance and complexity interconnection.

\begin{figure}[!htb]
	\centering
	\includegraphics[width=0.5\textwidth]{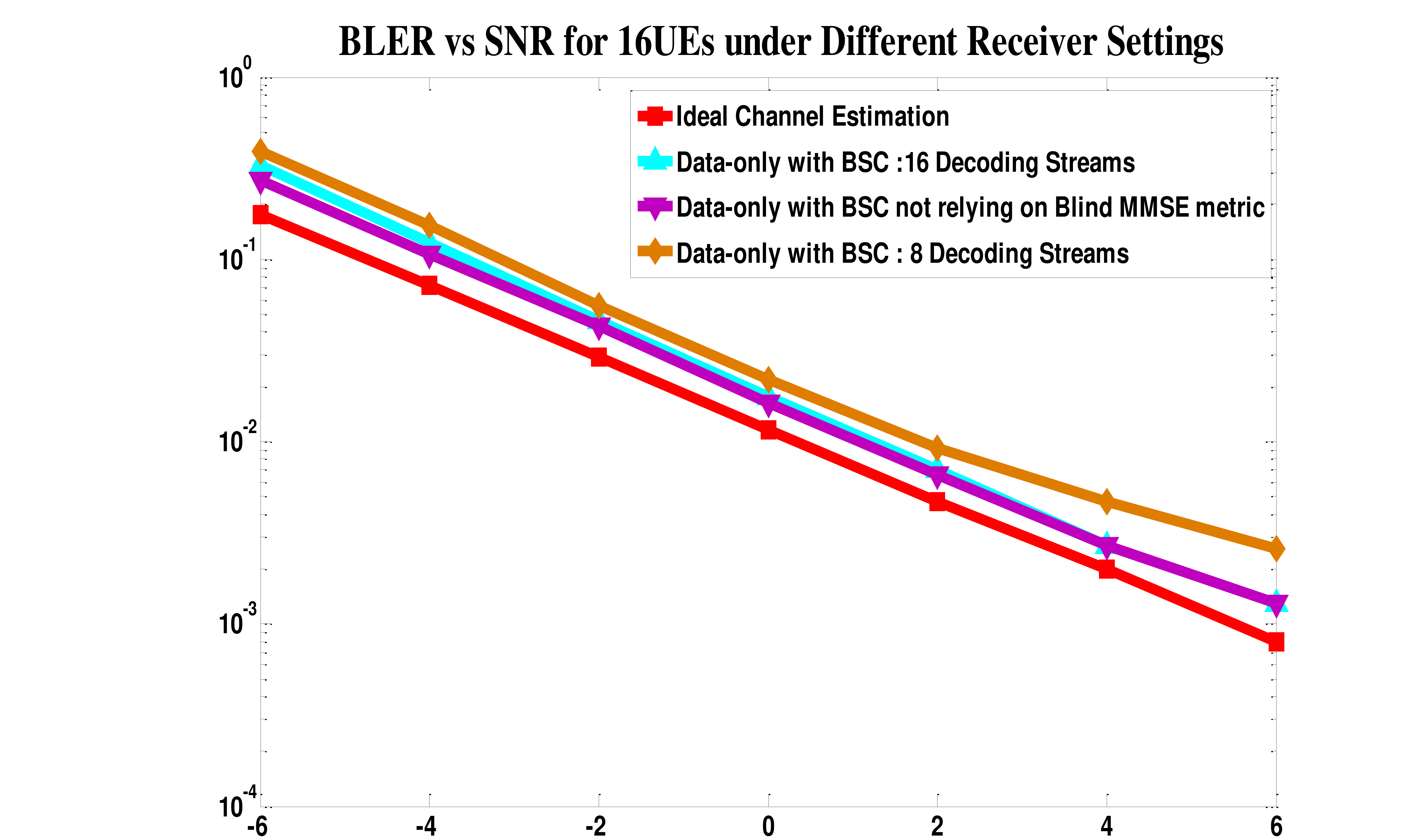}\\
	\caption{BLER vs SNR for Data-only Based Transmission of 16UEs}\label{blerEva16}
\end{figure}

\begin{figure}[!htb]
	\centering
	\includegraphics[width=0.5\textwidth]{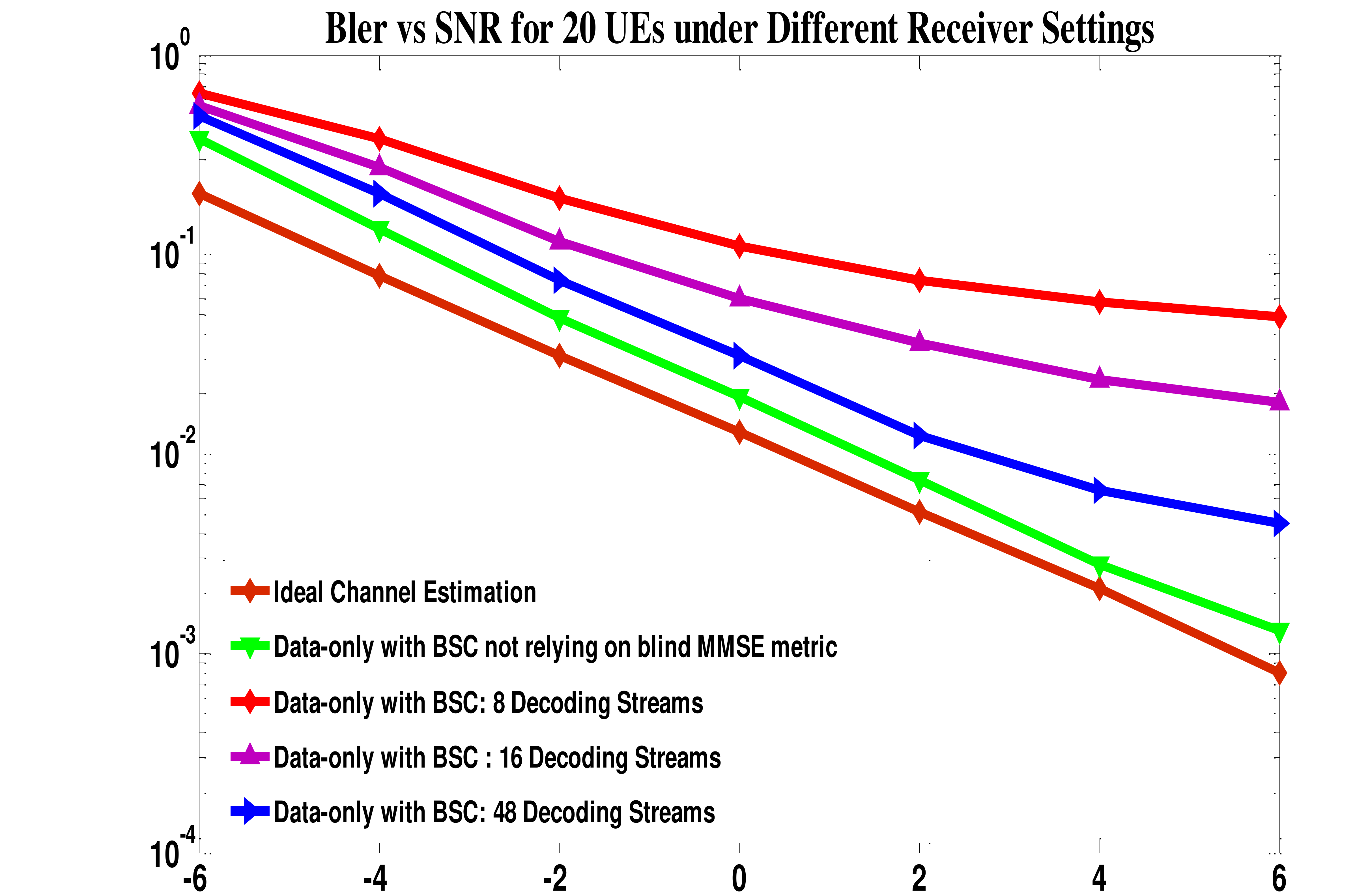}\\
	\caption{BLER vs SNR for Data-only AGF Transmission of 20 UEs}\label{blerEva20}
\end{figure}

Fig. \ref{blerEva16} illustrates that data-only MUD architecture could achieve 400\% overloading (16 simultaneous access UEs) at 2 $dB$ for BLER at $10^{-2}$ with MUSA spreading sequences at the transmitter accommodating all the receiver settings. The hierarchical blind UE detection mechanism adopting $M = 8$ and $F = 16$ could achieve similar performance to that of 384 times decoding efforts excluding any gradual down selection approach capable of considerable complexity reduction. For 500\% overloading (20 simultaneous access UEs), the BLER at 1\% could be achieved at around 3$dB$ for 48 decoding streams as shown in Fig. \ref{blerEva20}. It's worth noting that the proposed data-only MUD architecture could achieve close to ideal channel estimation performance. The ideal channel estimation evaluation assumes perfect channel estimation at each of the antennas and MMSE equalization in both the spatial and spreading code domain and the covariance matrix is therefore of dimension 8 in this case.

\begin{figure}[!htb]
	\centering
	\includegraphics[width=0.5\textwidth]{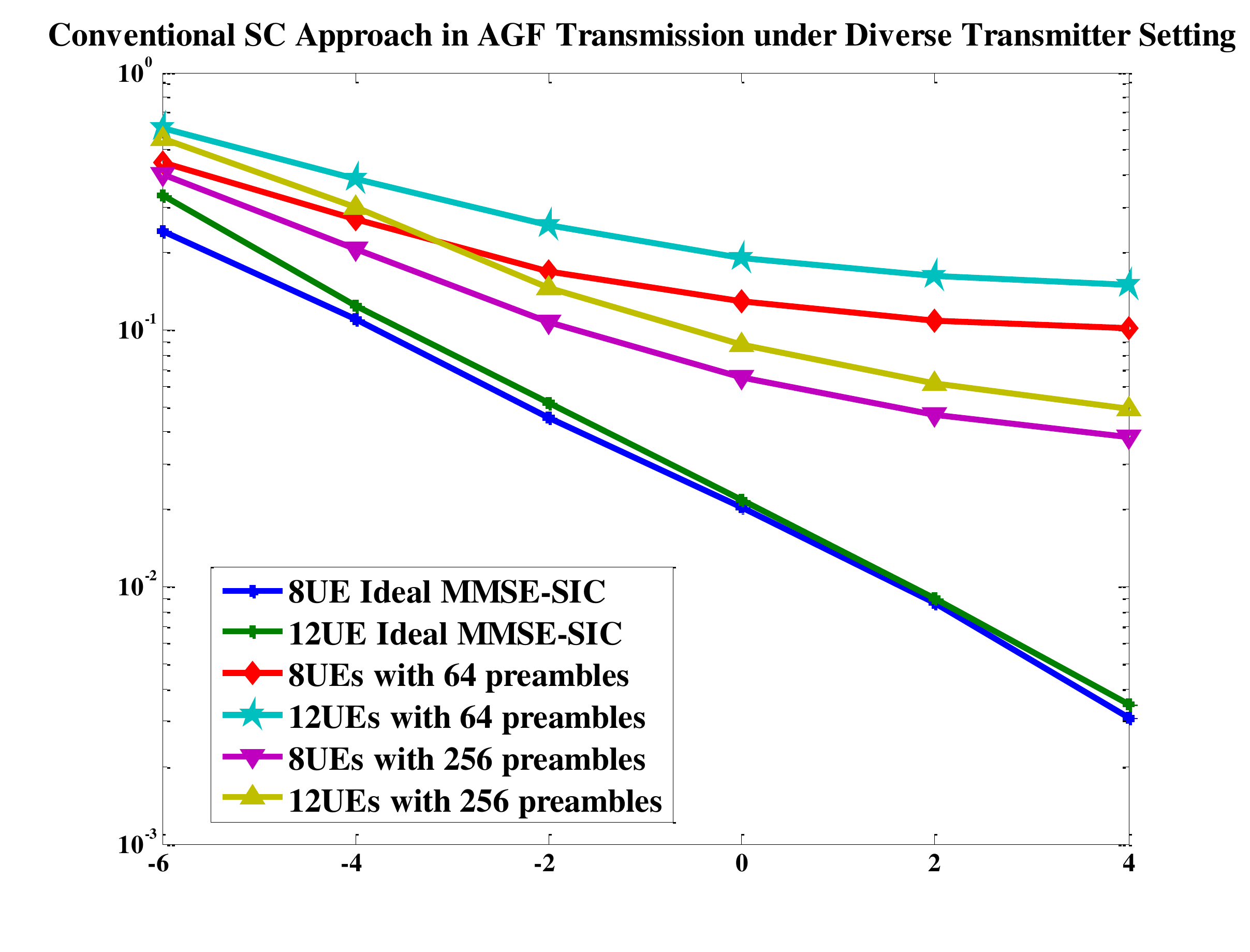}\\
	\caption{BLER vs SNR for Preamble Based AGF Transmission}\label{premP}
\end{figure}

BLER vs SNR performance adopting conventional spatial combining approach based on the channel estimated from the preambles are illustrated in Fig. \ref{premP} . The same amount of physical resources are utilized as that required by data-only transmission, half of which (2 resource blocks (RBs) consisting of 336 resource elements (REs)) are dedicated to carry the preambles.  64 and 256 orthogonal preambles are assumed for random selection of the 8 or 12 UEs respectively. In 12 UEs' case, the collision probability is over 30\% for the 64 preamble pool case \cite{ZTE.collision}, the size of which equals that of the MUSA sequence pool for random selection also. We could see non-negligible performance degradation of the preamble based blind MUD compared to that of both ideal channel estimation performance and data-only MUD.


\section{Conclusions}
This paper analyzes the design challenges as well as principles for AGF-HOL  transmission focusing in particular on blind spatial combining, following which state-of-art data-only MUD with blind spatial combining architecture are presented. While it would be difficult to perform accurate UE detection and channel estimation in the first place in the context of AGF-HOL transmission, it is indeed worthwhile to introduce the hierarchical mechanism capable of exploiting available information at different stages to achieve performance and complexity tradeoff to a balanced extent. Further research on blind activity detection metrics is necessary for the purpose of achieving lower overall blind MUD complexity in particular for the case with more receive antennas and handling the more realistic channel including To/Fo bias.

\bibliographystyle{IEEEtran}
\bibliography{HYZ_J}

\begin{thebibliography}{10}
\providecommand{\url}[1]{#1}
\csname url@samestyle\endcsname
\providecommand{\newblock}{\relax}
\providecommand{\bibinfo}[2]{#2}
\providecommand{\BIBentrySTDinterwordspacing}{\spaceskip=0pt\relax}
\providecommand{\BIBentryALTinterwordstretchfactor}{4}
\providecommand{\BIBentryALTinterwordspacing}{\spaceskip=\fontdimen2\font plus
\BIBentryALTinterwordstretchfactor\fontdimen3\font minus
  \fontdimen4\font\relax}
\providecommand{\BIBforeignlanguage}[2]{{%
\expandafter\ifx\csname l@#1\endcsname\relax
\typeout{** WARNING: IEEEtran.bst: No hyphenation pattern has been}%
\typeout{** loaded for the language `#1'. Using the pattern for}%
\typeout{** the default language instead.}%
\else
\language=\csname l@#1\endcsname
\fi
#2}}
\providecommand{\BIBdecl}{\relax}
\BIBdecl

\bibitem{3GPP.802}
{3GPP TR38.802}, ``Study on new radio access technology: Physical layer aspects
  (release 14),'' 2017.

\bibitem{Yuany.NOT}
Y.~Yuan, Z.~Yuan, G.~Yu, C.~h.~Hwang, P.~k.~Liao, A.~Li, and K.~Takeda,
  ``Non-orthogonal transmission technology in lte evolution,'' \emph{IEEE
  Communications Magazine}, 2016.

\bibitem{LDai.NomaSol}
L.~Dai, B.~Wang, Y.~Yuan, S.~Han, C.l.I, and Z.~Wang, ``Non-orthogonal multiple
  access for 5g: solutions, challenges, opportunities, and future research
  trends,'' \emph{IEEE Communications Magazine}, 2015.

\bibitem{bavand2017maximum}
M.Bavand and S.Blostein, ``Maximum signal minus interference to noise ratio
  multiuser receive beamforming,'' \emph{arXiv preprint arXiv:1705.05500},
  2017.

\bibitem{bavand2017user}
M.~Bavand and S.~D. Blostein, ``User selection and widely linear multiuser
  precoding for one-dimensional signalling,'' \emph{arXiv preprint
  arXiv:1705.09985}, 2017.

\bibitem{axiv}
Z.Yuan, Y.Hu, W.Li, and J.Dai, ``Blind multi-user detection for autonomous
  grant-free high-overloading {MA} without reference signal,''
  \emph{arXiv:1712.02601 [cs.IT]}, 2017.

\bibitem{zYuanvtc18}
Z.~Yuan, Y.~Hu, W.~Li, and J.~Dai, ``Blind multi-user detection for autonomous
  grant-free high-overloading multiple-access without reference signal,'' in
  \emph{IEEE 87th Vehicular Technology Spring Conference RAMAT}, 2018.

\bibitem{Yuan.BMU}
Z.~Yuan, C.~Yan, Y.~Yuan, and W.~Li, ``Blind multiple user detection for
  grant-free {MUSA} without reference signal,'' in \emph{IEEE 86th Vehicular
  Technology Conference}, 2017.

\bibitem{ZTErSINoMA}
Z.~M. ZTE, ``Rp-171043 revision of study on 5g non-orthogonal multiple
  access,'' 3GPP, Tech. Rep., 2017.

\bibitem{zhou2007iterative}
X.~Zhou, Z.~Shi, and M.~C. Reed, ``Iterative channel estimation for idma
  systems in time-varying channels,'' in \emph{Global Telecommunications
  Conference, 2007. GLOBECOM'07. IEEE}, 2007.

\bibitem{sergienko2017spectral}
A.~B. Sergienko and V.~P. Klimentyev, ``Spectral efficiency of uplink scma
  system with csi estimation,'' in \emph{Open Innovations Association (FRUCT),
  2017 20th Conference of}, 2017.

\bibitem{hammarberg2012channel}
P.~Hammarberg, F.~Rusek, and O.~Edfors, ``Channel estimation algorithms for
  ofdm-idma: complexity and performance,'' \emph{IEEE Transactions on Wireless
  Communications}, 2012.

\bibitem{ZTE.perf}
{3GPP R1-1608953}, ``Link-level performance evaluation for {MUSA},'' ZTE, ZTE
  Microelectronics.

\bibitem{yang2011receive}
J.~Yang, E.~Bjornson, and M.~Bengtsson, ``Receive beamforming design based on a
  multiple-state interference model,'' in \emph{Communications (ICC), 2011 IEEE
  International Conference on}, 2011.

\bibitem{gray1990vector}
R.~M. Gray, ``Vector quantization,'' in \emph{Readings in Speech Recognition},
  1990.

\bibitem{abdallah2014rate}
S.~Abdallah and S.~D. Blostein, ``Rate adaptation using long range channel
  prediction based on discrete prolate spheroidal sequences,'' in \emph{Signal
  Processing Advances in Wireless Communications (SPAWC), 2014 IEEE 15th
  International Workshop on}, 2014.

\bibitem{ZTE.collision}
{3GPP R1-1611500}, ``Further consideration on the preamble design for
  grant-free non-orthogonal {MA},'' ZTE, ZTE Microelectronics.

\end{thebibliography}
\end{document}